\documentclass[twocolumn,10pt]{IEEEtran}

\makeatletter
\def\ps@headings{%
\def\@oddhead{\mbox{}\scriptsize\rightmark \hfil \thepage}%
\def\@evenhead{\scriptsize\thepage \hfil \leftmark\mbox{}}%
\def\@oddfoot{}%
\def\@evenfoot{}}
\makeatother
\usepackage{amssymb}
\usepackage{amsmath}
\usepackage{balance}

\usepackage{amsthm}
\usepackage{amssymb}
\usepackage{graphicx,tikz}
\usetikzlibrary{arrows,automata}

\usepackage{array,cite}
\usepackage{subfigure,epsfig}

\title{\LARGE{Data Gathering in Networks of Bacteria Colonies:}\\ \LARGE{Collective Sensing and Relaying Using Molecular Communication} }
\author{ \normalsize Arash Einolghozati, Mohsen Sardari, Ahmad Beirami, Faramarz Fekri\\
School of Electrical and Computer Engineering, Georgia Institute of Technology, Atlanta, GA 30332\\
\texttt{Email:}\{einolghozati, mohsen.sardari, beirami, fekri\}@ece.gatech.edu
\thanks{This material is based upon work supported by the National Science Foundation under Grant No. CNS-111094}
}

\begin{document}

\maketitle
\thispagestyle{empty}
\pagestyle{empty}

\begin{abstract}
The prospect of new biological and industrial applications that require communication in micro-scale, encourages research on the design of bio-compatible communication networks using networking primitives already available in nature. One of the most promising candidates for constructing such networks is to adapt and engineer specific types of bacteria that are capable of sensing, actuation, and above all, communication with each other.
In this paper, we describe a new architecture for networks of bacteria to form
a data collecting network, as in traditional sensor networks. The key to this architecture is the fact that the node in the network itself is a bacterial colony; as an individual bacterium (biological agent) is a tiny unreliable element with limited capabilities. We describe such a network under two different scenarios. We study the data gathering (sensing and multihop communication) scenario as in sensor networks followed by the consensus problem in a multi-node network.
We will explain as to how the bacteria in the colony collectively orchestrate their actions as a node to perform sensing and relaying tasks that would not be possible (at least reliably) by an individual bacterium.
Each single bacterium in the colony forms a belief by sensing external parameter (e.g., a molecular signal from another node) from the medium and shares its belief with other bacteria in the colony. Then, after some interactions, all the bacteria in the colony form a common belief and act as a single node. We will model the reception process of each individual bacteria and will study its impact on the overall functionality of a node. We will present results on the reliability of the multihop communication for data gathering scenario as well as the speed of convergence in the consensus scenario.
\end{abstract}


\section{Introduction}
\label{sec:intro}

Emergence of new applications in biomedicine, e.g., smart drug administration and monitoring systems~\cite{cholesterol_nanosensor}, as well as several prospective industrial and security related applications including surveillance for biological and chemical attacks require network designs that are 1) capable of operation in micro-scale 2) bio-compatible, 3) cheap to deploy, 4) low energy and can harvest energy from the environment, and possibly 5) self-organized.
Conventional communication paradigms fail to satisfy all these requirements as deploying devices capable of generating electromagnetic signals in living organisms is inefficient, expensive, energy consuming, and can also be dangerous for the organism. Thus, such criteria necessitate the development of novel networking paradigms.
One of the most promising candidates for constructing networks that satisfy the requirements mentioned above is adaptation and engineering of specific types of bacteria that are capable of sensing, computation, actuation, and above all, communication with each other~\cite{Tamsir2011,Bassler1999}.
The existence of a form of communication using molecules that occurs naturally among bacteria has been confirmed. Communication enables single cells to gather and process sensory information about their environment and evaluate and react to chemical stimuli. Single cells use special types of molecules for sensing their environment or sending information to the outside world~\cite{Bassler1999}.


One phenomenon that in particular demonstrates the essential components of communication among cells is ``Quorum Sensing"~\cite{kaplan1985,Bassler1999,Hammer2003,Mehta2009}. Quorum Sensing can be viewed as a decentralized coordination process which allows bacteria to estimate the density of their population and regulate their behavior accordingly. To estimate the local population density, bacteria release specific signaling molecules. These molecules are subject to diffusion process that would make the molecules drift away instead of accumulating in the bacteria vicinity~\cite{Bassler1999, muller2008,perez2010}. These signal molecules will then reach the neighboring bacteria providing them with information about other bacteria in the environment.
 As the local density of bacteria increases, so will the concentration of the molecules in the medium. Bacteria have molecule receptors that can estimate the molecular concentration and thus the bacteria population density. Bacteria use quorum sensing to coordinate actions (mostly energy expensive) that cannot be carried out by a single bacterium.
This process, captures most of the important components of a communication system in micro-scale.

Arguably, the most dominating form of communication at the scale of microorganisms, as observed in communication among bacteria, is molecular communication, and in particular, Diffusion based Molecular Communication (DbMC)~\cite{Akyildiz2011,ISIT2011_Arash,ITW2011_Arash,Pierobon2010}. The essence of DbMC is embedding the information in the alteration of the concentration of the molecules and relying on diffusion to transfer the information. Hence, the behavior of a single molecule which exhibits so much randomness is of no importance. Understanding primitives of information processing and communication in bacteria would enable us to design networks with far more capabilities in their domain.

\begin{figure*}
\vspace{-0.05in}
\begin{center}
  \subfigure[A hypothetical network of biological nodes.]{
  \includegraphics[width=.18\textwidth,angle=-90]{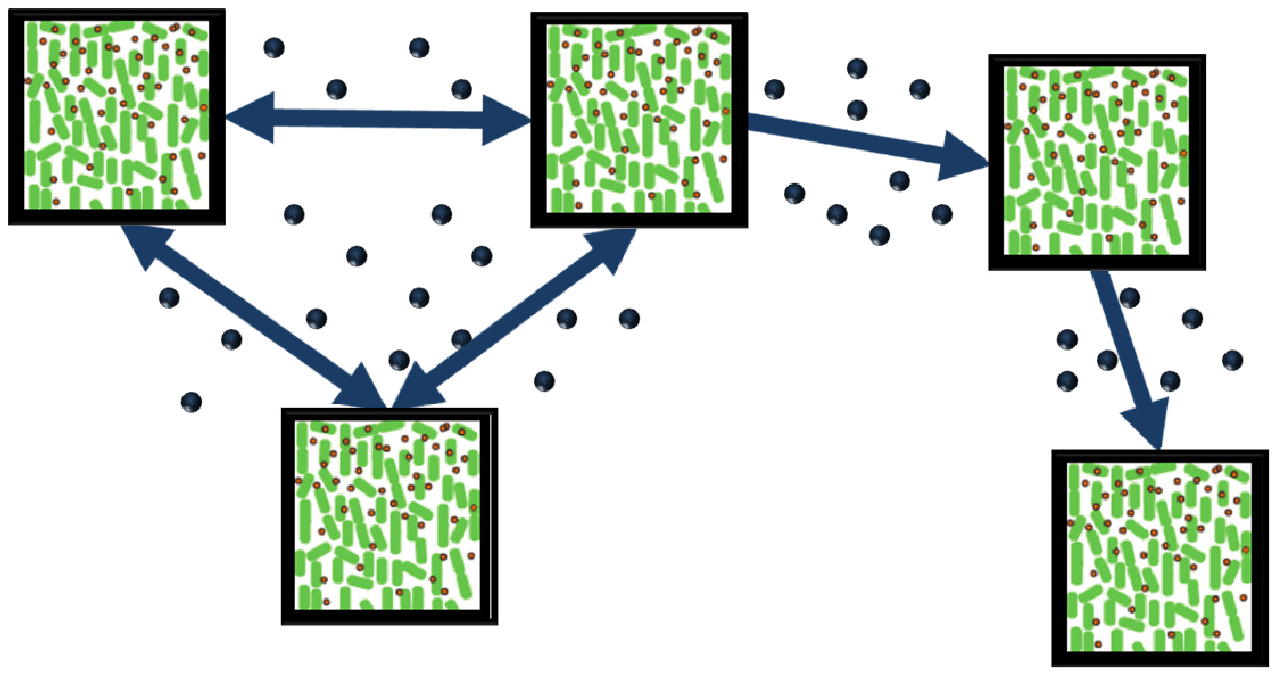}
  \label{fig:network-model}
  }
  \hspace{0.05\textwidth}
  \subfigure[A biological Node.]{
  \includegraphics[width=.18\textwidth,angle=-90]{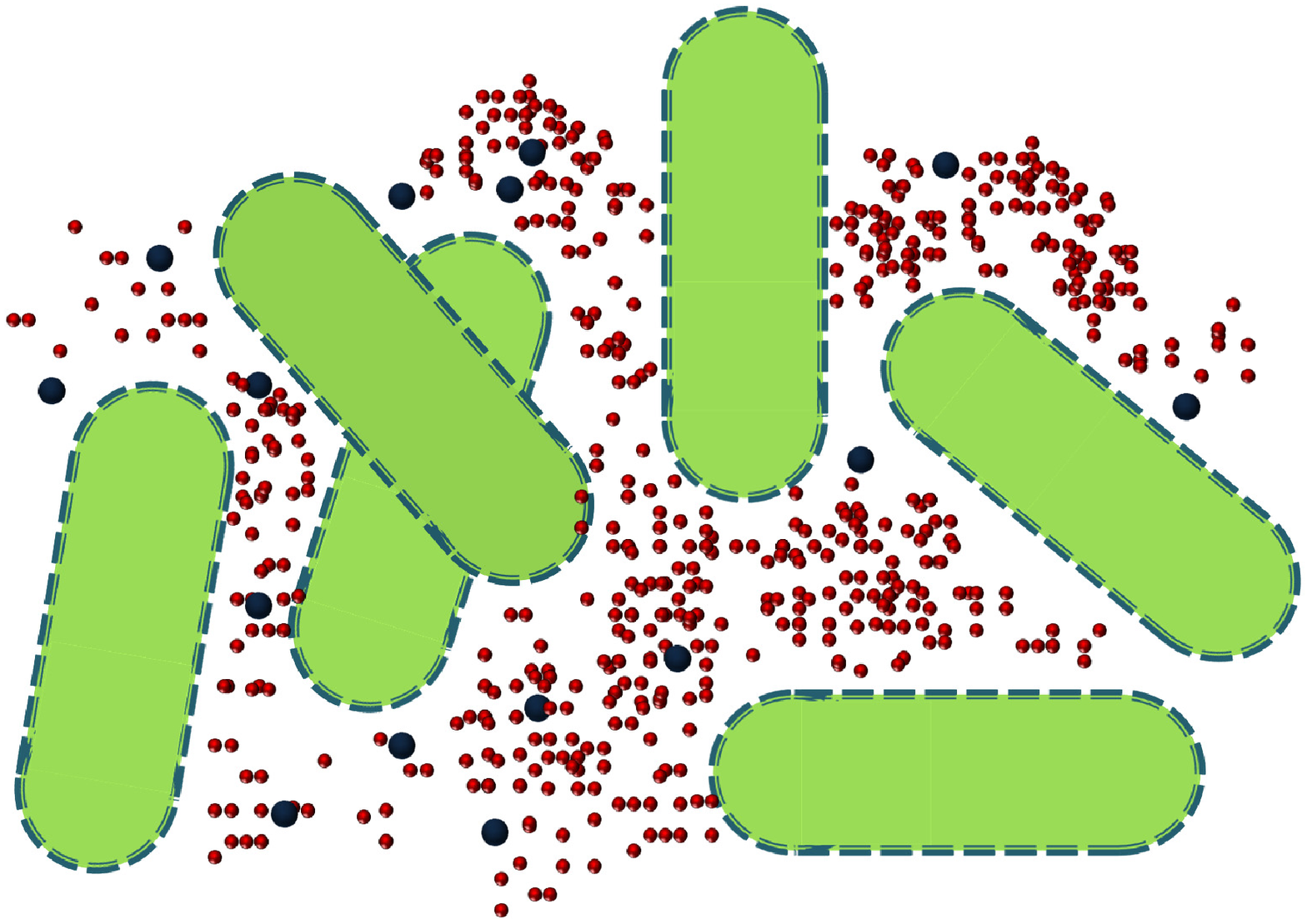}
  \label{fig:supernode}
  }
  \hspace{0.05\textwidth}
  \subfigure[A biological agent.]{
  \includegraphics[width=.18\textwidth,angle=-90]{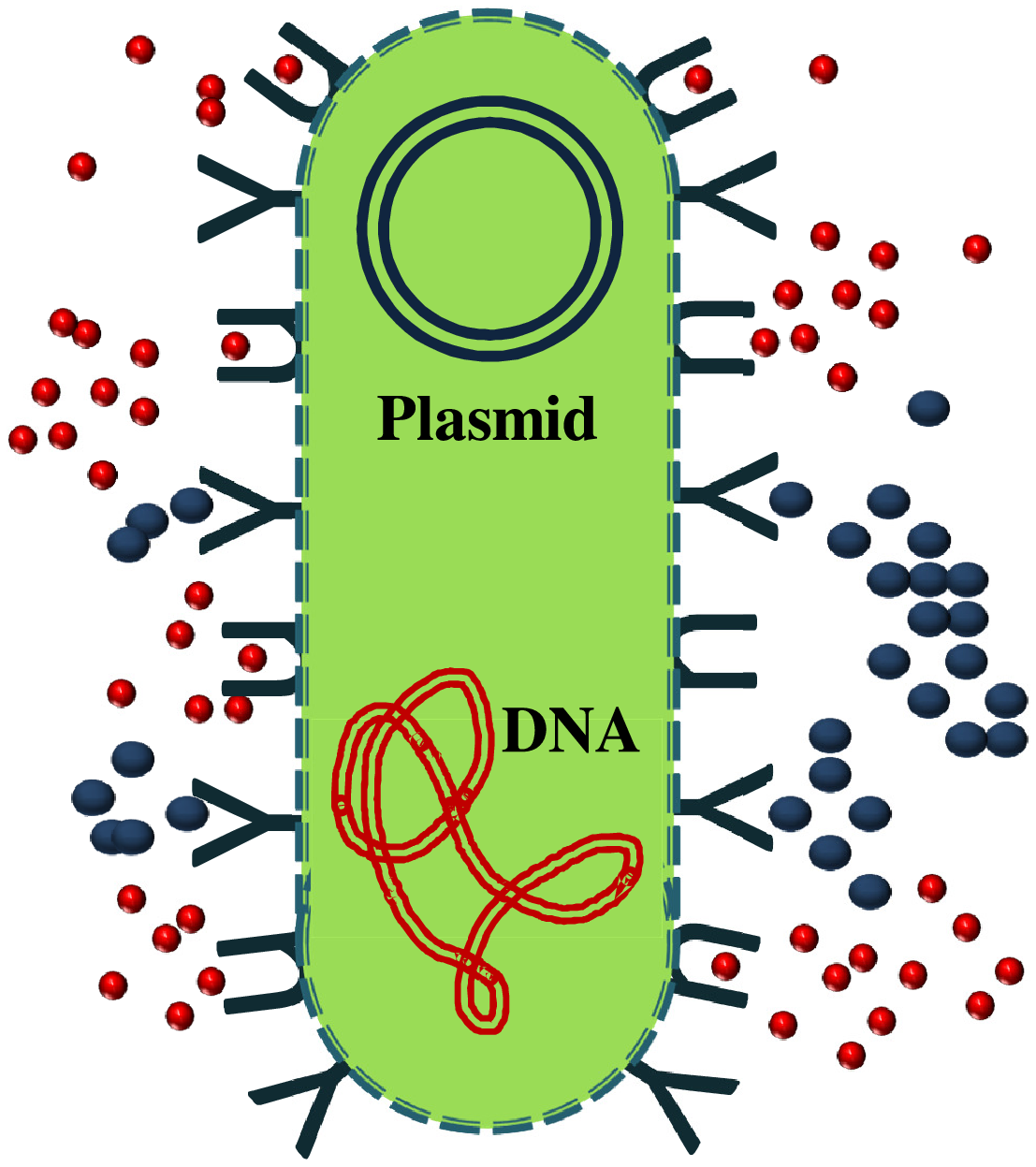}
  \label{fig:agent-model}
  }
\end{center}
\caption{Configuration and constituent modules of MCN.}
\label{fig:network_model}
\end{figure*}

In our previous works, we studied some of the relevant molecular communication problems, especially, the impact of diffusion channel~\cite{ISIT2011_Arash} and molecular receptors~\cite{ITW2011_Arash} on the achievable rates of communication between a source and a destination.
In this paper, inspired from the natural molecular communication and collective sensing occurring in bacteria, we focus on the networking and propose the Molecular Communication Networking (MCN). MCN utilizes the primitives of molecular communication to engineer a network of bacterial colonies capable of data gathering and relaying.

In contrast to the previous vision that the molecular communication incurs between two entities (e.g. two bacteria), here, we focus on an architecture in which a cluster of biological entities (i.e. a cluster of bacteria) communicate with another cluster. We will refer to these clusters as biological \emph{nodes}. The basic building block of the network, hereafter called as the \emph{agent}, are genetically modified bacteria (or primitive biological elements that have simple capabilities similar to those of bacteria) to transfer information from one point to another or alternatively implement a sensor network. These biological entities are very primitive and unreliable, and hence, incapable of providing reliable communication and sensing by themselves. However, when a cluster of these biological agents form a biological node, they are collectively capable of reliable transmission and reception of information.
The information to be transferred is acquired from monitoring some features of the environment by the nodes. Agents in a node collaboratively form a belief regarding the environment (collective sensing) and act as a whole. The information may be binary such as whether or not a specific chemical substance is present in the environment or it can be the quantity of a chemical substance or even temperature in the surrounding environment.
The interplay of these individual agents that form the biological nodes capable of collective sensing and relaying information is studied in this paper.


The rest of the paper is organized as follows.
In Sec.~\ref{sec:config}, we describe the constituent modules in a typical molecular communication network, namely the agents, nodes, and inter-node communication. In Sec.~\ref{sec:agent}, we describe the capabilities and roles of the agents in the network. In Sec.~\ref{sec:intra}, we describe the collective sensing and communication of agents in a node. In Sec.~\ref{sec:inter}, we describe the inter-node communication paradigms that we envision in the MCN and Sec.~\ref{sec:conclusion} concludes the paper.

\section{The Building Blocks of Molecular Communication Networks }
\label{sec:config}

In this section, we describe the configuration and constituent modules of MCN that we study.
The schematic for a general MCN is shown in Fig.~\ref{fig:network_model}. The basic elements of MCN are agents that are engineered bacteria. The placement of these agents and their interconnectivity play a key role in forming the network. In Fig.~\ref{fig:network-model} we have depicted a hypothetical network where a number of nodes (to be defined later) on the left perform a form of distributed processing whose result is then relayed to a sink node on the right. The very basic questions that arise are 1) what constitutes a node, and 2) how is a message transmitted between the nodes.

Although a single biological agent is capable of those tasks by itself, it would be very challenging (if not impossible) to form a reliable network out of these agents each acting as a single node (such as a single node in sensor networks). The rational is that a single agent (which is going to be a single cell) is too primitive to carry the transmission and reception tasks with high reliability. It is more reasonable to cluster these agents and orchestrate the cluster activity to perform these sophisticated tasks (e.g., sensing/reception and transmission). Therefore, we envision that a number of biological agents form a cluster that we refer to as a \emph{node}, depicted in Fig.~\ref{fig:supernode}. One realization of our vision of a node is the trapping chamber implemented on a chip~\cite{Danino2010,Prindle2011}. All the agents in a node are able to communicate with each other directly. The molecule production of each agent affects the concentration sensed by the other agents in the node. On the other hand, this strategy enables these primitive agents to collaboratively produce high concentration of molecules that can travel and be sensed from longer distances, i.e., the overall response of molecule production of several agents will be the superposition of individual responses.
In other words, \emph{nodes in MCN are a population of genetically engineered bacteria}.

The notion of the node introduced above encourages investigation of two modes of communication in the network: \emph{intra-node collective sensing} and \emph{inter-node communication}. The intra-node collective sensing is the process by which agents in a node aggregate their sensing and coordinate their actions for inter-node communication. The intra-node collective sensing can be thought of as a modified form of quorum sensing process performed naturally. On the other hand, inter-node communication is used for transferring the information of one node to another node which can be farther apart. It is important to note that intra-node and inter-node communication processes can be decoupled by considering a different type of molecule for each communication. For example, in-node processing and coordination can be performed with the molecules of type I and inter-node communication can be independently performed with the molecules of type II.
 The idea of using multiple molecules for the molecular communication is inspired by the presence of quorum sensing processes with multiple molecule types in some bacteria families such as Vibrio Harveyi. This bacterium is believed to use multiple molecule types for quorum sensing; one type can only be detected by Vibrio Harveyi bacteria while others can be detected by the entire Vibrio family~\cite{Bassler_Harveyi}.

In short, functionalities of a hypothetical MCN network are as follows. First, in intra-node collective sensing phase, all the agents in a node measure the environment for a parameter of interest such as existence of a chemical substance or even incoming molecules from another node. Then, they share that information with one another via intra-node communication. In the second phase, the node communicates with the neighboring nodes and transfer its information, i.e, inter-node communication. These two modes of communication will be discussed in detail in the following sections.
%
%
%
\section{Agent Model}
\label{sec:agent}
Agents are the primitive elements of our network that can be realized by genetic modification of bacteria. Parts of DNA content of a bacterium can be manipulated to construct specific features like signal transmission and/or reception. One common technique involves manipulation of DNA molecules within bacteria that are separate from the chromosomal DNA called \emph{plasmid}~\cite{Lipps2008,Cobo2010}, as shown in Fig.~\ref{fig:agent-model}. A plasmid is a double-stranded DNA molecule that can be viewed as an ordering of genes each responsible for a function. For example, deactivating the receptor (transmitter) genes in some variants of bacteria can provide transmitter (receiver) agents.

Agents can also work as transceivers, i.e., they sense the concentration of molecules in the medium and release molecules at a specific rate back into the medium. More sophisticated designs can also be implemented through the same process. For example, some agents can be engineered to be able to receive specific types of molecules and perform particular logical operations upon reception~\cite{Tamsir2011}.

 Here, we focus on the molecule reception (sensing) process. Assume each agent has $N$ identical receptors for a specific molecule type, for example, the U-shaped receptors in Fig.~\ref{fig:agent-model}. As in~\cite{ISIT2011_Arash,ITW2011_Arash}, the information is embedded in the concentration of the molecules. In steady-state the probability $p$ that a molecule is trapped in a receptor is directly related to the concentration of molecules $\rho$ at the vicinity of receptors. By estimating $p$, the agent obtains an estimate for the concentration $\rho$ and hence, can receive and decode the information.
To each receptor, we associate an indicator random variable $X_i$, where $i \in \left\{1,2,\ldots,N\right\}$, which shows whether a molecule is trapped in that receptor. Here, $X_i$\rq{}s are Bernoulli distributed with parameter $p$, i.e., $\mathbf{P}(X_i=1)=p$. It is known that $\sum_{i=1}^N X_i$ is the sufficient statistics for estimation of $p$. This implies that the receiver functionality is to add up the values of all receptors to obtain the best estimate for $p$. This counting process introduces a random noise to the reception. Let $\hat{p}$ be the best unbiased estimator for $p$. Since $\hat{p}=\frac{1}{N}\sum_{i=1}^N X_i$, the expected value and variance of $\hat{p}$, given $p$, are $\mathbf{E}[\hat{p}]=p$ and $\text{Var}[\hat{p}]=\frac{1}{N}p(1-p)$.
Thus, $\hat{p}$ depends on the value of ${p}$ as
\begin{equation}
\label{eq:conditional}
\mathbf{P}\left( \hat{p} \left. = \frac{i}{N}\right|p\right)={N\choose i} p^i (1-p)^ {N-i},\; i \in \left\{0,1,\ldots,N\right\}
\end{equation}
Hence, each agent has $N+1$ different outputs that occur with different probabilities. In~\cite{ITW2011_Arash}, we obtained the distribution on $p$ which results in the largest mutual information between $p$ and $\hat{p}$ which is Arcsine distribution:
$$
f_P(p)=\frac{1}{\pi\sqrt{p(1-p)}}, \quad 0<p<1.
$$
From this distribution and the one-to-one function that relates $\rho$ to $p$, we can design the signaling (i.e. the concentration levels) at the transmitter node for the multi-level modulation. The above model would enable us to have a simple understanding of the impact of different components of the agent, e.g., the molecule production circuitry, on the overall functionality of the node.

Next, we study the characteristic of the nodes. As will be discussed in detail in the next section, upon sensing the molecule concentration (receiving a signal), each agent forms a local belief. Then, the agents in the node share their beliefs via intra-node communication to reach unanimity. The end of the collective sensing phase will then trigger the inter-node communication. For example, a colony of bacteria can be engineered such that the collective sensing inside the colony triggers the genes to produce the type II molecules into the environment and start the inter-node communication. 
Each node can be considered as an independent entity which is able to sense the concentration of different types of molecules, process the information and release molecues.
In order to send molecules at an arbitrary rate, the node should produce a stimulus to arise an appropriate number of agents. This stimulus activates a number of the agents (which depends on the intensity of the stimulus) and the agents produce molecules with constant rate. The large number of agents enables the node to send molecules with practically arbitrary rates.


\section{Intra-Node Collective Sensing}
\label{sec:intra}

Thus far, we introduced the agents and briefly described the envisioned capabilities of the nodes in the network. In this section, we study the collective sensing process inside a node. In the sequel, we first give a brief review about the general communication model for two agents using DbMC in the medium and extend the idea to intra-node communication, i.e., communication between all the agents within a node. Then, we describe how agents communicate with each other to form a unanimous belief within a node.

\vspace{-.1in}
\subsection{DbMC Model}

Molecular communication is based on diffusion~\cite{Pierobon2010}.
Molecules are produced by the transmitter agents and are injected into the channel, which is the medium that carries them toward the receiver.
The propagation of molecules in the medium follows the Fick's second law of diffusion which states that the concentration $c(x,t)$ of molecules at position $x$ at time $t$ in an $m$-dimensional space, when the molecule production rate at the source is $r(x,t)$, is~\cite{random_walk}
%
\begin{equation}
\label{eq:eq4}
 \frac {\partial c(x,t)}{\partial t} = D \nabla^2 c(x,t)+r(x,t),
\end{equation}
where, $D$ is the diffusion coefficient of the medium. The impulse response of~(\ref{eq:eq4}) is denoted as the Green's function~\cite{random_walk}.
%
%
Let $c^*(x,t)$ be the impulse response for a 3-D medium for an arbitrary input $F(t)$ at source, which can be calculated by convolving the input rate $F(t)$ with the impulse response. Hence, we will have~\cite{random_walk}
%
%
%
\begin{equation}
\label{eq:diffusion}
c^*(x,t) =\int^{\infty}_0 F(\tau)\frac {1}{(4\pi D(t-\tau))^{\frac{3}{2}}} \exp{\left( -\frac{x^2}{4D(t-\tau)}\right)} \,d{\tau}.
\end{equation}


In our setup, agents transmit molecules with constant rates in the time. Hence $F(t)=\alpha u(t)$ where $u(t)$ is the step function and $\alpha$ is the amplitude of the input rate. The steady-state of the response in~(\ref{eq:diffusion}) is the concentration that agents sense. The integral in~(\ref{eq:diffusion}) is bounded and can be calculated for each $x$. Therefore, the steady-state response will be $\alpha G(x)$ where $G$ is a function depending only on the distance from the transmitter and the diffusion coefficient. Since the diffusion equation is linear, the steady-state response due to the molecule production of the multiple transmitting agents would be $c^*(x)=\sum_i \alpha_i G(d_i)$, where $d_i$ is the distance of agent $i$ to the location $x$ and $\alpha_i$ is its respective production rate.
\vspace{-.05in}
\subsection{Belief Formation at the Node: Intra-Node Communication}
As stated previously, a node must sense a parameter or receive and decode a signal. In the belief formation phase, agents of the node employ the type I molecules to communicate among themselves and reach a unanimous estimate of the sensed parameter or molecular signal. This phase of communication is inspired by Quorum Sensing (QS) in bacteria. By QS, the information about the number of bacteria reaching a threshold or not is propagated in the network of bacteria. Hence, bacteria can reach an unanimous estimate about this binary parameter by producing and sensing molecules at the same time. Here, we consider sensing a binary parameter in the environment in which agents should reach consensus.

The belief formation can be studied in two setups: with or without the feedback from agent's receptors to its transmitters. In the first case, which is similar to bacteria in the nature, the agents increase the rate of production of molecules if they sense a higher concentration in the environment than what they had previously estimated. Without the feedback, the agents are programmed to produce molecules based solely on their own estimate and this rate remains constant until the steady state. The latter scenario can be realized in bacteria by removing the genes responsible for the positive feedback.The first approach, i.e., with feedback, ensures faster convergence in the network but could lead to a false alarm when the incorrect estimate is magnified due to the feedback.

Here, we explain how the scenario without feedback can work. We assume that $n$ agents in a node are sufficiently nearby each other that every agent can hear every other agents in the node. Each agent is able to sense the binary parameter of interest and become activated with probability $p_0$ which depends on the intensity of the parameter and sensitivity of the sensors of each agent. Upon activation, each agent produces molecules with rate $r_0$ or otherwise does nothing. In a compact group of agents, we can assume that the agents sense the same concentration in steady state which is the average of concentration around them and hence, there is an average distance $d_0$ for all the agents in~(\ref{eq:diffusion}). Therefore, the steady-state concentration of molecules will be $c^*=n_{on} r_0 G(d_0) $ where $n_{on}$ is the number of active agents and $G(\cdot)$ is the integral in~(\ref{eq:diffusion}). Hence, by sensing the concentration of molecules in their vicinity, agents can estimate the number of active nodes.  Finally, comparing it to a threshold determines whether the parameter of interest is ON or OFF in the environment. The above discussion can be easily extended to the case where in the parameter of interest or the signal concentration takes multi-level values. This extension can be done by considering discrete levels for the parameter value and sending molecules with rates corresponding to those levels.

\section{Inter-Node Communication in MCN}
\label{sec:inter}
Unlike the intra-node communication, the inter-node communication is more general and not limited to collective sensing between the agents. Nodes can be considered as independent and relatively smart entities in a network and can use communication in the network for variety of reasons. We study two important communication scenarios in the network. Note that the application of intra-node communication is not limited to these two scenarios. First, we consider a relaying scenario where information is transferred from a source node to a destination node several hops away. The relaying problem is in particular of importance because of the very low communication range in MCN. The concentration of molecules decays exponentially with respect to distance from the transmitter. In other words, nodes can only directly communicate with each other in short distances. Therefore, for applications that require transfer of information for long distances, we should resort to multihop communication strategy. In the second scenario, we study consensus problem in a multinode network where each node performs measurement of a parameter of interest and shares its measurement data with other nodes.


\vspace{-.05in}
\subsection{Relay Problem in MCN}

\begin{figure}
\vspace{-0.8in}
\hspace{-.3in}
\includegraphics[height=1.15\linewidth,angle=-90]{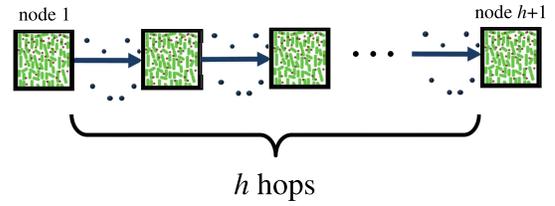}
\vspace{-1.2in}
\caption{Multihop information relaying}
\label{fig:relay_model}
\end{figure}

Consider node $1$ shown in Fig.~\ref{fig:relay_model}. Node 1 has some information that needs to be transmitted to node $h+1$. This information is an estimate of a local parameter that has been obtained via the belief formation by the agents at node $1$, i.e., the source node. We assume that nodes $1$ and $h+1$ are located outside the effective communication range, and hence, they should use other nodes as relays to transfer information. Suppose that the effective range of each node includes only the immediate neighbors of it.

In a general network,  it remains as an open research problem as to how one can perform the path discovery in a molecular communication while maintaining the realistic assumptions on the biological nodes (e.g., limited number of molecules to be released, lack of even simple addressing method, etc.). To avoid path discovery issues, we envision that all paths from the nodes to the sink node can be preformed for the sensor network applications. This is realistic as the data is collected from the nodes to the fixed sink node and hence, the path from each node to the sink can be formed by the proper placement of the nodes.
Specifically, to form a path (as in a routing path in the traditional networks), we assume that every node has two forms of type II molecules (call it type II-A and type II-B). A node can receive the signal on one type (e.g.,  type II-A) and transmit the
signal on the other type. The placement of the nodes on the path (and the entire network) must ensure that the molecule type that a node transmits to its neighbor node (or nodes) matches the neighbor's receiving molecule type.

There is a total number of $h$ hops needed to transfer the information from node $1$ to $h+1$. We consider the communication between node $i$ and $i+1$ in the path. Due to the significant distance of the nodes relative to the size of the agents, we can assume all agents are co-located at the center of the node and sense the same concentration of molecules.
Note that the processes of production of molecules, diffusion in the channel, and the sensing impose significant delay, which is in order of several hours. Further, the lingering of the molecules in the medium (before they are carried away) makes the channel to have \emph{memory}, which in turn causes inter symbol interference~\cite{ISIT2011_Arash}. Hence, in the molecular communication, we try to transmit as many bits of information as possible in one use of channel.
We consider agents to have binary outputs, i.e., they either produce molecules with a specific concentration rate or they are off. Hence, given $n$ agents per node, there are potentially $n+1$ possible levels of concentrations at the steady state. This is equivalent to $\log_2 {(n+1)}$ bits of information that can be transferred to node $i+1$.
Each agent at node $i+1$ senses the concentration received in the steady-state. As a result, the node forms an estimate by using the belief formation mechanism explained in Sec. \ref{sec:intra}. Based on the received level of concentration, node $i+1$ activates a fraction of its agents to produce type II molecules to be sent to the next hop. 

Because of the probabilistic behavior of the signal reception by the agents, information is relayed one hop at a time with some error. Two main factors influence the  reliability of communication in each hop, namely the number of agents in the node and the number of receptors in each agent. The number of agents $n$ in the node determines the range of possible detectable inputs. A node with $n$ agents is capable of receiving and decoding $n+1$ distinct levels of the parameter of the interest (recall that the parameter of interest could be the level of chemical substance, or the concentration of the type II molecular signal received at the node). On the other hand, the number of receptors $N$ of each agent determines the precision by which the concentration can be sensed.

\begin{figure}
\centering
\vspace{0.05in}
\includegraphics[width =.98\linewidth]{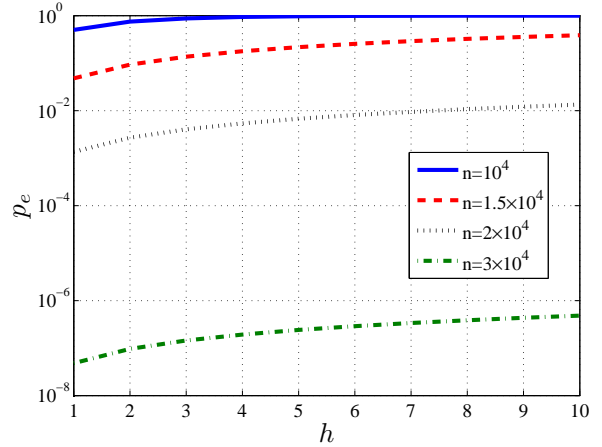}
\vspace{-.05in}
\caption{Probability of error in relaying versus number of hops for different values of the number of agents in each node $n$ .}
\vspace{-0.1in}
\label{fig:error_hop}
\end{figure}

In order to further illustrate the problem of reliability, we consider the case of relaying a binary measurement (e.g., existence of a chemical substance) from node $1$ to $h+1$. In the ON state, all the agents in node $i$ produce molecules with a specific rate $r_0$. Hence, the steady-state concentration at node $i+1$ at the distance $d_0$ will be $c^*=n r_0 G(d_0)$. Let each receptor of an agent be activated with probability $p_0$, which is a function of $c^*$. Each agent is activated if at least a threshold $k$ out of the $N$ of its receptors are activated. Hence, $p_a$ the probability of the activation of an agent is given by $p_a=\text{Pr}\left(B(N,p_0)\geq k\right)$, where $B(N,p_0)$ denotes the binomial distribution. Thus, the expected number of activated agents is $E(n_a)=n p_a$. Node $i+1$ decides whether or not to stimulate its agents to produce molecules based on its belief, which is formed through the collective sensing process described in Sec.~\ref{sec:intra}.  In this setting, an error will occur if the number of activated agents in the final node $h+1$ is smaller than the threshold for it to turn on. The aggregate probability of error is obtained by accumulating the probability of error in all the hops. The end-to-end probability of error in a typical network  is plotted in Fig.~\ref{fig:error_hop} for different number of agents in a node versus the number of hops. As shown in the plot, increasing the number of agents results in higher range of reliable relaying.

\vspace{-.05in}
\subsection{Consensus Problem in MCN}

We now consider the consensus problem in a sensor network formed by MCN. In a general consensus problem, nodes in a network communicate with each other to obtain the best estimate given their initial estimates. The average value of these initial estimates is considered to be an important measure that can be considered as a goal in a network. Here, by producing different rates of molecules, nodes give information about their own estimates and by sensing the concentration of molecules in their proximity, they obtain information about the estimates of other nodes. The key idea here is the linear nature of diffusion which lets nodes to determine the aggregation of estimates of other agents.

\begin{figure}
\centering
\vspace{0.05in}
\includegraphics[width =.98\linewidth]{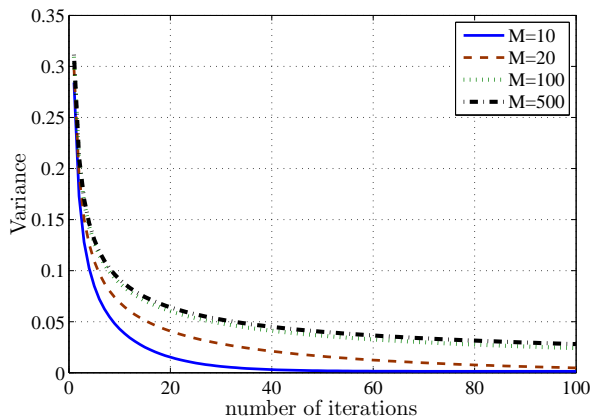}
\vspace{-0.16in}
\caption{Convergence of the iterative algorithm versus the number of nodes in the network.}
\vspace{-0.14in}
\label{fig:converge}
\end{figure}

In~\cite{CISS2011_Arash}, we introduced an algorithm that results in consensus in an extended network. We consider a network of $M$ nodes deployed uniformly. We assume that the initial estimates of the environment parameter $\theta$ by nodes has Gaussian distribution with mean $\mu$ and variance $\sigma^2_0$. The average of initial estimates is the best unbiased estimate for $\mu$ with variance $\frac{\sigma_0^2}{M}$. In a network with sufficiently large $M$, each node observes the same relative distances $d_{ij}$ to the other nodes in the network (except for the nodes on the boundary of the network whose effects are negligible when $M$ is large). Since the network size is large, each node has effective communications with only a specific number $M^{\prime}$ of nodes which is determined by the range of nodes. By using the diffusion equation, the concentration of molecules in the steady-state can be derived by
$
{\bf c^*}={\bf G} {\bf r},
$
where ${\bf c^*}$ and ${\bf r}$ are both vectors and ${\bf G}$ is a symmetric matrix in which ${\bf G}(i,j)$ is equal to $G(d_{ij})$  if two nodes have effective communication with each other or zero otherwise. In addition, columns and rows of $\bf G$ are permutations of each other.

The proposed algorithm in~\cite{CISS2011_Arash} is an iterative algorithm such that in iteration $l$, node $i$ sends molecules with rate $\frac{\hat{\theta}_i(l)}{S}$ where $S$ is the sum of a column of $\bf G$ which is the same for the all columns and $\hat{\theta}_i(l)$ is the estimate of node $i$ at iteration $l$. Then, it updates its estimate as $\hat{\theta}_i(l+1)=c^*(l)$. Note that the process of sensing the concentration, updating the estimates and sending with new rates take some time that we can assume the channel has been rest meanwhile. In~\cite{CISS2011_Arash}, We showed that due to properties of the matrix $G$, this iterative algorithm will converge to $\mu$ the average of initial beliefs. Moreover, by performing more iterations, the variance of beliefs of nodes can become arbitrarily close to its lower bound $\frac{\sigma_0^2}{M}$ which approaches to zero in a network with large number of nodes.

It can be proved that when matrix ${\bf G}$ becomes more sparse, i.e. the columns and the rows contain more zeros, the rate of convergence falls. This can be explained by the fact that a more sparse ${\bf G}$ means a more extended network. Hence, more number of iterations are needed to spread the information in the network and hence, reaching consensus will be more time consuming. In particular when ${\bf G}$ becomes a diagonal matrix, the variance does not converge to $\frac{\sigma_0^2}{M}$. This case is equivalent to the scenario in which none of the nodes are able to communicate with each other and their initial beliefs cannot be improved. The simulation results for convergence in a uniform network are shown in Fig.~\ref{fig:converge} for different values of $M$. As we observe in the plot, the initial variance of estimates decreases with number of iterations and the convergence occurs sooner for smaller networks.

\section{conclusion}
\label{sec:conclusion}
In this paper, we studied data gathering in the networks of bacteria colonies. We proposed that these networks can be formed via  several bacteria agents acting as a whole like a node; performing collective sensing, signal reception and transmission. Two types of communication, one between agents of a nodes and the other between nodes in a network were studied. We demonstrated that the beliefs of the nodes can be formed and communicated reliably via the orchestrated action of the tiny unreliable agents in a node. Such networks are envisioned to perform tasks in biological environments where conventional networking paradigms fail to operate.

%
%
%
%

\bibliographystyle{IEEEtran}
\bibliography{InfoCom2011}
\end{document}